\documentclass[12pt]{article}

\usepackage{graphicx}

\begin{document}
\begin{center}
{\bf On exponential modified gravity}\\
\vspace{5mm}
 S. I. Kruglov \\
\vspace{5mm}

\textit{Department of Chemical and Physical Sciences, University of Toronto,\\
3359 Mississauga Rd. North, Mississauga, Ontario, Canada L5L 1C6}
\footnote{E-mail: krouglov@utsc.utoronto.ca}
\end{center}

\vspace{5mm}
\begin{abstract}
A modified theory of gravity with the function $F(R) = R\exp\left(\alpha R\right)$ instead of Ricci scalar $R$ in the Einstein$-$Hilbert action is considered and analyzed. The action of the model is converted into Einstein$-$Hilbert action at small value of the parameter $\alpha$. From local tests we obtain a bound on the parameter $\alpha\leq 10^{-6}$ cm$^2$. The Jordan and Einstein frames are investigated and the potential of the scalar field in Einstein's frame is found. The mass of a scalar degree of freedom as a function of curvature is obtained. The static solutions of the model are found corresponding to the Schwarzschild$-$de Sitter space. We show that the de Sitter space is unstable but a solution with zero curvature is stable. The cosmological parameters of the model are calculated. It was demonstrated that the model passes the matter stability test.
\end{abstract}
\vspace{5mm}

\section{Introduction}

Astronomical data indicate that the Universe accelerates at the present time. The nature of the driving force that results the accelerated expansion is unknown yet. There are several approaches to explain cosmic acceleration. The interpretation of currently accelerating Universe, within the General Relativity (GR), requires the introduction of dark energy (DE) - exotic substance with large negative pressure $P_{DE}$ so that $P_{DE}\simeq -\rho_{DE}$ ($\rho_{DE}$ is the dark energy density). The similar scheme uses the cosmological constant $\Lambda$ \cite{Frieman}. Such a model gives a good description of all observational data. However, with theoretical point of view, it is not clear how to explain the introduction of a new physical constant $\Lambda$ which is very small compared with vacuum energy of elementary particles. Models with dynamical dark energy include a new scalar field \cite{Linde}. Another way to describe the acceleration of the early and late Universe is the modification of GR. So-called $F(R)$-gravity theories replace the Ricci scalar in Einstein$-$Hilbert action by the function $F(R)$ \cite{Caldwell}, \cite{Faraoni}, \cite{Capozziello}, \cite{Odintsov}. It should be noted that the first successful examples of viable DE description in F(R)-gravity were given in \cite{Hu}, \cite{Appl}, \cite{Star}.
Such purely gravitational models present an alternative to $\Lambda$CDM ($\Lambda$-Cold Dark Matter) model and may clear up the coincidence problem, and describe the inflation and late-time acceleration. In $F(R)$-gravity models the cosmic acceleration is due to new gravitational physics.

In this paper a model of exponential-like $F(R)$-gravity is considered and investigated.
The paper is organized as follows. In Sec.2, we consider a model of modified gravity with the exponential-like Lagrangian density. A bound on the parameter $\alpha$ with the dimension
(length)$^2$ is obtained. We find static Schwarzschild$-$de Sitter solutions in Jordan's frame and describe FRW (Friedmann$-$Robertson$-$Walker) cosmology in Sec.3. The potential of the scalar field in the scalar-tensor form of the model (in Einstein's frame) is obtained in Sec.4. It is shown that the de Sitter space is unstable and the Minkowski space corresponding to a solution with zero Ricci scalar is stable. In Sec.5 the matter stability of the model is investigated and we demonstrate that the model passes the matter stability test at $R_0<2/\alpha$. The cosmological parameters of the model are calculated in Sec.6. In Sec.7 we discuss results obtained.

The Minkowski metric $\eta_{\mu\nu}$=diag(-1, 1, 1, 1) is used and  $c=\hbar=1$ is assumed throughout the paper.

\section{The modified gravity model}

We consider the modified gravitational theory with the function $F(R)$ instead of the Ricci curvature $R$ ($R\rightarrow F(R)$) in the Einstein$-$Hilbert action:
\begin{equation}
F(R)=R\exp\left(\alpha R\right),
\label{1}
\end{equation}
so that the action is
\begin{equation}
S=\int d^4x\sqrt{-g}{\cal L}=\int d^4x\sqrt{-g}\left[\frac{1}{2\kappa^2}F(R)+{\cal L}_m\right],
\label{2}
\end{equation}
where $\kappa=\sqrt{8\pi}m_{Pl}^{-1}$, $g$=det$g_{\alpha \beta}$ ($g_{\alpha \beta}$ is a metric tensor), $m_{Pl}=G^{-1/2}$ is the Planck mass, $G$ is the gravitation (Newton) constant,
${\cal L}_m$ is the matter Lagrangian density. Thus, the constant $\alpha$ with the dimension of (length)$^2$ is introduced. The action (2) is written in the Jordan frame.
It should be mentioned that some variants of exponential gravity were considered in \cite{Amendola}, \cite{Carloni}, \cite{Carloni1}, \cite{Linder}. To pass the Solar System tests the constant $\alpha$ should be small compared with $R^{-1}$ ($\alpha R\ll 1$) because the deviation from GR based on the Einstein$-$Hilbert action has to be diminutive.
As a result one can obtain from (1) the Taylor series
\begin{equation}
F(R)=R+\alpha R^2+\frac{1}{2}\alpha^2 R^3 +....
\label{3}
\end{equation}
AS $\lim_{\alpha\rightarrow 0}F(R)=R$, action (2) of our model approaches to the Einstein $-$Hilbert action at $\alpha R\ll 1$.
GR passes local tests and, therefore, one may obtain a restriction on the parameter $\alpha$ from observational data. From the  E\"{o}t-Wash experiment \cite{Kapner}, \cite{Jetzer} (see also \cite{Berry}, \cite{Zhuk}), we obtain a laboratory bound on the parameter $\alpha$:
\begin{equation}
\alpha\leq 10^{-6} cm^2.
\label{4}
\end{equation}
It should be mentioned that F(R)-gravity with the function $F(R)=R+R^2/6M^2$ ($M$ has a dimension of the mass) was considered by Starobinsky \cite{Starobinsky} which is the approximation to series (3) at small $\alpha R$ (the value $\alpha R$ is dimensionless). The modified $R^2$-gravity is insensitively investigated \cite{Appleby}, \cite{Arbuzova}.  Quantum corrections to GR include $R^2$ counter term \cite{Stelle}, \cite{Birrel} as well as Ricci tensor squared ($R_{\mu\nu}R^{\mu\nu}$) which includes ghosts. Therefore, for any form of the function $F(R)$, $F(R)$-gravity is not renormalizable. Nevertheless, such models possess attractive features: an absence singularity in the past and in the future and give the self-consistent inflation, etc.

Let us consider the matter Lagrangian density ${\cal L}_m$, entering (2), which represents the perfect fluid with the energy-momentum tensor
\begin{equation}
T^{mat}_{\alpha\beta}= \left(P^{mat}+\rho^{mat}\right)u_\alpha u_\beta+P^{mat}g_{\alpha\beta},
\label{5}
\end{equation}
where $P^{mat}$ is a pressure, $\rho^{mat}$ is the energy density, and the four-velocity of the fluid obeys $u^\alpha u_\alpha=-1$. Then equations of motion following from Eq.(2) are given by
\begin{equation}
R_{\mu\nu}F'(R)-\frac{1}{2}g_{\mu\nu}F(R)+g_{\mu\nu}g^{\alpha\beta}\nabla_\alpha\nabla_\beta F'(R) -\nabla_\mu\nabla_\nu F'(R)=\kappa^2T^{mat}_{\mu\nu},
\label{6}
\end{equation}
where a covariant derivative is $\nabla_\mu$, $F'(R)=dF(R)/dR$. For FRW metric, the conservation of the energy-momentum tensor $\nabla^\mu T^{mat}_{\mu\nu} = 0$ results:
\begin{equation}
\dot{\rho}^{mat} + 3H\left(\rho^{mat} + P^{mat}\right)=0.
\label{7}
\end{equation}
Here the Hubble parameter is $H = \dot{a}(t)/a(t)$, where a(t) is a scale factor and a over dot denotes the differentiation with respect to the time. It follows from Eq.(7) that for the fluid with the property of the dark energy when the equation of state (EoS) is $P^{mat}=-\rho^{mat}$, the energy density $\rho^{mat}$ is a constant.

\section{Static Solutions}

Let us consider solutions to Eq. (6) for action (2) in a case with a constant Ricci scalar $R=R_0$ without any matter. Then Eq. (6) reads \cite{Barrow}
\begin{equation}
2F(R_0)-R_0F'(R_0)=0.
\label{8}
\end{equation}
From Eq.(8), we obtain the equation as follows:
\begin{equation}
R_0\left(1-\alpha R_0\right)=0.
\label{9}
\end{equation}
There are two solutions to Eq. (9):
\begin{equation}
 R_0=0,~~~~~~ R_0=\frac{1}{\alpha}.
\label{10}
\end{equation}
The conditions of classical and quantum stability \cite{Appleby} $F'(R)>0$, $F''(R)>0$ are realized in our model for $\alpha>0$, $R>0$.
As a result, both solutions lead to the Schwarzschild$-$de Sitter space. Non-trivial solution (10) $R_0=1/\alpha$ corresponds to early-time inflation. The solution with vanishing curvature $R_0=0$ leads to the Minkowski space. If $F'(R_0)/F''(R_0)>R_0$, the positive solution may describe primordial and present dark energy which is future stable \cite{Schmidt}. For the function (1), we obtain
\begin{equation}
\frac{F'(R_0)}{F''(R_0)}=\frac{1+\alpha R}{\alpha\left(2+\alpha R\right)}.
\label{11}
\end{equation}
For the Minkowski space-time, $R_0=0$, and $F'(R_0)/F''(R_0)=1/(2\alpha)>0$. Therefore, the flat space-time is stable. The static solution $R_0=1/\alpha$, corresponding to the de Sitter space-time, is unstable because $F'(R_0)/F''(R_0)=2/(3\alpha)<R_0$.
Of course the concrete scenario of the Universe evolution can be described after finding the exact solutions to Eq.(6) for Ricci scalar depending on the time.

The Schwarzschild spherically symmetric metric is given by
\begin{equation}
ds^2=-B(r)dt^2+\frac{dr^2}{B(r)}+r^2\left(d\theta^2+\sin^2\theta d\phi^2\right).
\label{12}
\end{equation}
For the constant Ricci scalar $R_0$, $F(R)$-gravity theories possess
Schwarzschild$-$ (anti-)de Sitter solutions with the function $B(r)$:
\begin{equation}
B(r)=1-\frac{2MG}{r}-\frac{R_0}{12}r^2,
\label{13}
\end{equation}
with the mass of the black hole $M$. For $R_0>0$ the de Sitter space is realized, and the case $R_0<0$
corresponds to the anti-de Sitter space. As in our model the non-trivial solution (10) is $R_0=1/\alpha >0$, it corresponds to the de Sitter space and the function (13) is
\begin{equation}
B(r)=1-\frac{2MG}{r}-\frac{1}{12\alpha}r^2.
\label{14}
\end{equation}
Because $\alpha >0$ we have the classical stability of Schwarzschild black holes. Solutions of Einstein's equation with cosmological constant $\Lambda$ have also the function of the form (13) with $R_0=4\Lambda$. Therefore, the model under consideration leads to the dynamical cosmological constant $\Lambda=1/(4\alpha)$ at the time when $R=R_0=1/\alpha$. Thus, even for the space without any matter, the model mimics the dark energy (the cosmological constant). The similar property of other $F(R)$ models was discussed in \cite{Barrow}, \cite{Carroll}.

In $F(R)$-gravity the entropy $S$ is given as follows \cite{Akbar}, \cite{Gong}, \cite{Brustein}:
\begin{equation}
S =\frac{F'(R)A}{4G},
\label{15}
\end{equation}
with the area of the horizon $A$. Eq.(15) is the generalization of the Bekenstein-Hawking formula \cite{Bekenstein}, \cite{Hawking} on the case of $F(R)$-gravity. One obtains from equation (1) $F'(R)=(1+\alpha R)\exp(\alpha R)$, and
entropy (15) becomes
\begin{equation}
S =\frac{\left(1+\alpha R\right)\exp\left(\alpha R\right)A}{4G}.
\label{16}
\end{equation}
As a result, one can introduce the effective gravitational coupling $G_{eff}=G/(1+\alpha R)\exp(\alpha R)$. The nontrivial solution (10) $R_0=1/\alpha$ gives the effective gravitational constant $G_{eff}=G/(2e)$ at the time of inflation when $R=R_0=1/\alpha$.

\subsection{FRW cosmology}

The homogeneous, isotropic and spatially flat FRW cosmology is described by
the space-time metric
\begin{equation}
ds^2 = -dt^2 + a^2(t)\left(dx^2+dy^2+dz^2\right).
\label{17}
\end{equation}
In this case the Ricci scalar R is given by $R = 12H^2 + 6\dot{H}$. Then Eq.(6) can be represented as follows:
\begin{equation}
\frac{F(R)}{2}-3\left(H^2+\dot{H}\right)F'(R)+18\left(4H^2\dot{H}+ H\ddot{H}\right)F''(R)=\kappa^2\rho^{(m)},
\label{18}
\end{equation}
\[
\frac{F(R)}{2}-\left(3H^2+\dot{H}\right)F'(R)+6\left(8H^2\dot{H}+4\dot{H}^2+6H\ddot{H}+
\partial_t\ddot{H}\right)F''(R)
\]
\vspace{-7mm}
\begin{equation}
\label{19}
\end{equation}
\vspace{-7mm}
\[
+36\left(4H\dot{H}+\ddot{H}\right)F'''(R)=-\kappa^2P^{(m)}.
\]
For solutions $H_0$=const, $\dot{H}_0=0$, Eqs.(18), (19) are consistent with EoS of dark energy $\rho^{(m)}=-P^{(m)}$. The model without matter is considered here. Then $H_0=\sqrt{R_0/12}$ and we obtain from Eq.(10) for a de Sitter phase $H_0=1/\sqrt{12\alpha}$, and a scale factor becomes
\begin{equation}
a(t)=a_0\exp\left(\frac{t}{2\sqrt{3\alpha}}\right),
\label{20}
\end{equation}
where $a(0)$ is a scale factor at a cosmic time $t=0$. Solution (20) describes the the eternal inflation phase, i.e. has no end. To describe inflation in details, one needs to obtain the exact solution for the general case $R\neq \textrm{const}$.

\section{The Scalar-Tensor Form}

\begin{figure}[h]
\includegraphics[height=4.0in,width=4.0in]{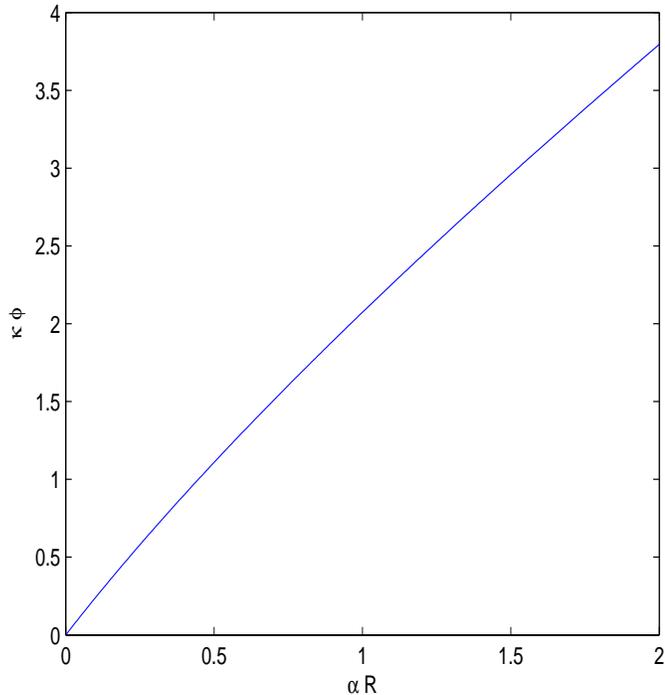}
\caption{\label{fig.1}$\kappa\phi$ versus $\alpha R$.}
\end{figure}

\begin{figure}[h]
\includegraphics[height=4.0in,width=4.0in]{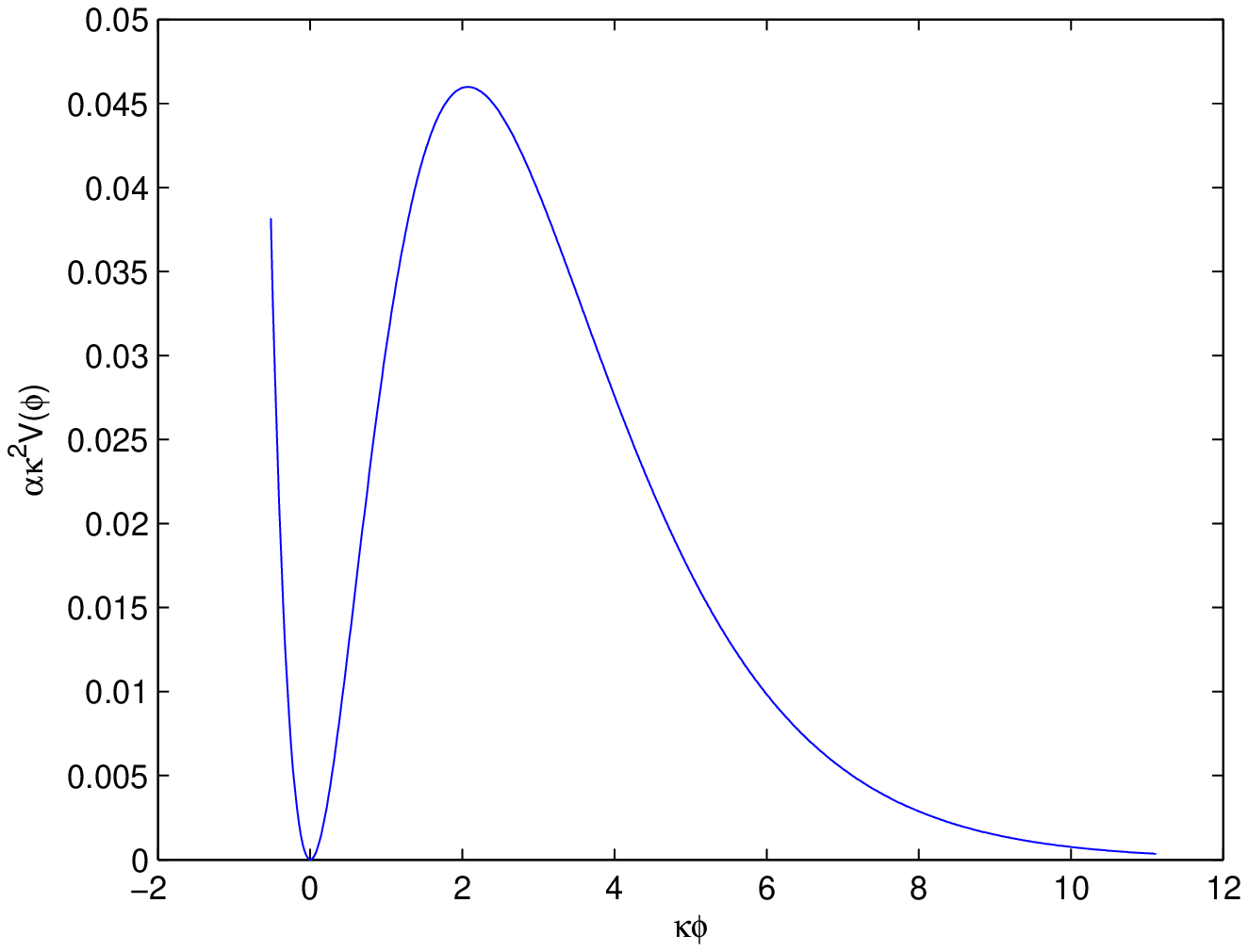}
\caption{\label{fig.2}$\alpha\kappa^2 V(\phi)$ versus $\kappa\phi$.}
\end{figure}

\begin{figure}[h]
\includegraphics[height=4.0in,width=4.0in]{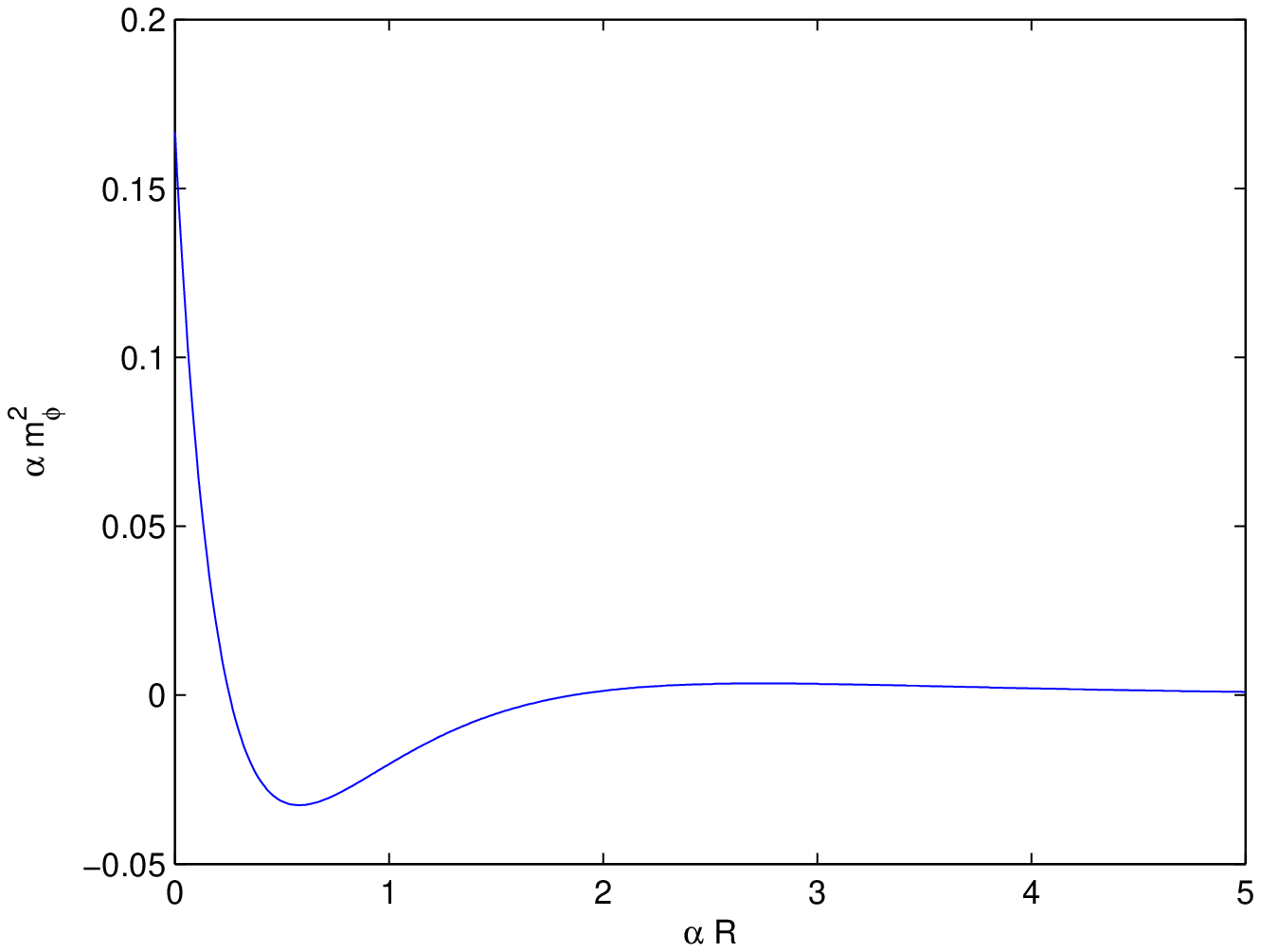}
\caption{\label{fig.3}$\alpha m^2_\phi $ versus $\alpha R$.}
\end{figure}

We have formulated modified $F(R)$-gravity in the Jordan frame with tensor variables $g_{\mu\nu}$. Another description, in the Einstein frame, corresponds to the scalar-tensor theory of gravity with conformally transformed metric \cite{Sokolowski}
\begin{equation}
\widetilde{g}_{\mu\nu} =F'(R)g_{\mu\nu}=\left(1+\alpha R)\exp(\alpha R\right)g_{\mu\nu}.
\label{21}
\end{equation}
In new variables, equation (2) for ${\cal L}_m=0$, in the Einstein frame, is given by
\begin{equation}
S=\int d^4x\sqrt{-g}\left[\frac{1}{2\kappa^2}\widetilde{R}-\frac{1}{2}\widetilde{g}^{\mu\nu}
\nabla_\mu\phi\nabla_\nu\phi-V(\phi)\right],
\label{22}
\end{equation}
where $\widetilde{R}$ is defined by new metric (21). The scalar field $\phi$ and the potential $V(\phi)$ are given by equations
\begin{equation}
\phi=\frac{\sqrt{3}\ln F'(R)}{\sqrt{2}\kappa}=\frac{\sqrt{3}}{\sqrt{2}\kappa}\left[\ln \left(1+\alpha R\right)+\alpha R\right],
\label{23}
\end{equation}
\begin{equation}
V(\phi) =\frac{RF'(R)-F(R)}{2\kappa^2F'^2(R)}|_{R=R(\phi)}=\frac{\alpha R^2\exp\left(-\alpha R\right)}{2\kappa^2\left(1+\alpha R\right)^2}|_{R=R(\phi)},
\label{24}
\end{equation}
where the scalar curvature $R$ in Eq.(24) being the solution of transcendental Eq.(23), $R(\phi)$. The graph of the function $\kappa\phi (R)$ is given in Fig.\ref{fig.1}.
On can verify that the potential (24) has minimum at $R_0=0$ ($V'=0$, $V''>0$) and maximum at $R_0=1/\alpha$ ($V'=0$, $V''<0$). Thus, the flat space-time (zero scalar curvature) is the stable state of the Universe and the state with $R_0=1/\alpha$ is unstable.
The plot of the function $V(\phi)$ (24) is presented in Fig.\ref{fig.2}.
It should be mentioned that in the Einstein frame free particles of matter do not move in space-time geodesics because of interactions with the scalar field $\phi$. There is a correction in the right hand side of the geodesic equation representing a fifth force. Because the fifth force depends on space-time (and proportional to $\nabla^\mu \phi$) the universality of free fall (Weak Equivalence Principle) is violated. The mass squared of a scalar state is defined by the equation \cite{Faraoni}
\begin{equation}
m_\phi^2=\frac{d^2V}{d\phi^2} =\frac{1}{3}\left(\frac{1}{F''(R)}+\frac{R}{F'(R)}-\frac{4F(R)}{F'^2(R)}\right).
\label{25}
\end{equation}
We obtain from Eq.(1) the mass squared of a scalar field
\begin{equation}
m_\phi^2=\frac{\left[1-4\alpha R +\left(\alpha R\right)^3\right]\exp\left(-\alpha R\right)}
{3\alpha\left(2+\alpha R\right)\left(1+\alpha R\right)^2}.
\label{26}
\end{equation}
For $R_0=0$ the value $m_\phi^2=1/(6\alpha)$ is positive and the solution with zero scalar curvature, corresponding to the Minkowski space, gives a stabile state. The de Sitter solution (10)
$R_0=1/\alpha$ leads to the negative mass squared $m_\phi^2=-1/(18e\alpha)$. Thus, again we conclude that the de Sitter space with $R_0=1/\alpha$ is unstable. One can verify from Eq.(26) that at $\alpha R<0.25$ $m_\phi^2>0$, i.e. the space-time is stable. The graph of the function (26) $m^2_\phi$ is given in Fig.\ref{fig.3}.
The criterion of the stability of the de Sitter solution in F(R) gravity was first obtained in \cite{Schmidt}. For small value of $\alpha$ corrections to the Newton law are negligible. One can say that after the Big Bang the Universe is in unstable de Sitter's phase and inflates (rapidly expands) having the positive curvature $R_0=1/\alpha$. Then the curvature decreases and the Universe approaches to the stable Minkowski space.

\section{Matter Stability}

From equation (6), after taking the trace, one obtains the equation of motion for a curvature scalar
\begin{equation}
3g^{\alpha\beta}\nabla_\alpha\nabla_\beta F'(R)+ F'(R)R-2F(R)=\kappa^2T^{mat},
\label{27}
\end{equation}
where $T^{mat}=T^{mat}_{\mu\nu}g^{\mu\nu} $. Following \cite{Dolgov}, to investigate the matter stability, we consider Eq.(27) for weak gravity objects. For a flat Minkowski metric and spatially constant distribution Eq.(27) reads
\begin{equation}
-3F^{(2)}(R)\ddot{R}-3F^{(3)}(R)\dot{R}^2+ F^{(1)}(R)R-2F(R)=\kappa^2T^{mat},
\label{28}
\end{equation}
with the notation $F^{(n)}(R)=d^nF(R)/dR^n$. From Eq.(1), we find
\begin{equation}
F^{(n)}(R)=\alpha^{n-1}\left(n+\alpha R\right)\exp\left(\alpha R\right).
\label{29}
\end{equation}
Let us consider a perturbation so that $R=R_0+R_1$, $|R_1|\ll|R_0|$, and according to GR $R_0 =- \kappa^2T^{mat}$. Then Eq.(28) leads to (see \cite{Dolgov}, \cite{Odintsov})
\[
\ddot{R}_0+\ddot{R}_1+\frac{F^{(3)}(R_0)}{F^{(2)}(R_0)}\left(\dot{R}_0^2+2\dot{R}_0\dot{R}_1\right)
\]
\vspace{-7mm}
\begin{equation} \label{30}
\end{equation}
\vspace{-7mm}
\[
+ \frac{2F(R_0)-R_0\left[1+F^{(1)}(R_0)\right]}{3F^{(2)}(R_0)}=U(R_0)R_1,
\]
with
\[
U(R_0)=\frac{F^{(3)2}-F^{(2)}F^{(4)}}{F^{(2)2}}\dot{R_0}^2
\]
\vspace{-7mm}
\begin{equation} \label{31}
\end{equation}
\vspace{-7mm}
\[
+ \frac{\left(R_0F^{(2)}-F^{(1)}\right)F^{(2)}+\left(2F-R_0F^{(1)}-R_0\right)
F^{(3)}}{3F^{(2)2}}.
\]
For $U(R_0)>0$, $R_1$ exponentially increases in the time and the system is unstable. From Eqs.(29), (31), we obtain
\begin{equation}
U(R_0)=\frac{\left(\alpha \dot{R}_0\right)^2}{\left(2+\alpha R_0\right)^2}
+ \frac{\alpha R_0-2}{3\alpha \left(2+\alpha R_0\right)^2}.
\label{32}
\end{equation}
Implying that the rate $(\alpha \dot{R}_0)^2$ is small compared to $1/\alpha$, $(\alpha \dot{R}_0)^2\ll 1/\alpha$, for a matter stability $U(R_0)<0$, one comes to the condition
\begin{equation}
R_0<\frac{2}{\alpha}.
\label{33}
\end{equation}
For static solutions (10) the inequality (33) is satisfied and the model passes the matter stability test. Eq.(33) gives the restriction on the biggest curvature of the Universe. One can introduce the fundamental length $L=\sqrt{\alpha}$ so that the smallest size of the Universe is $L$. It should be mentioned that the Born$-$Infeld-like modified gravity model \cite{Kruglov} possesses the similar features.

\section{Cosmological parameters of the model}

\begin{figure}[h]
\includegraphics[height=4.0in,width=4.0in]{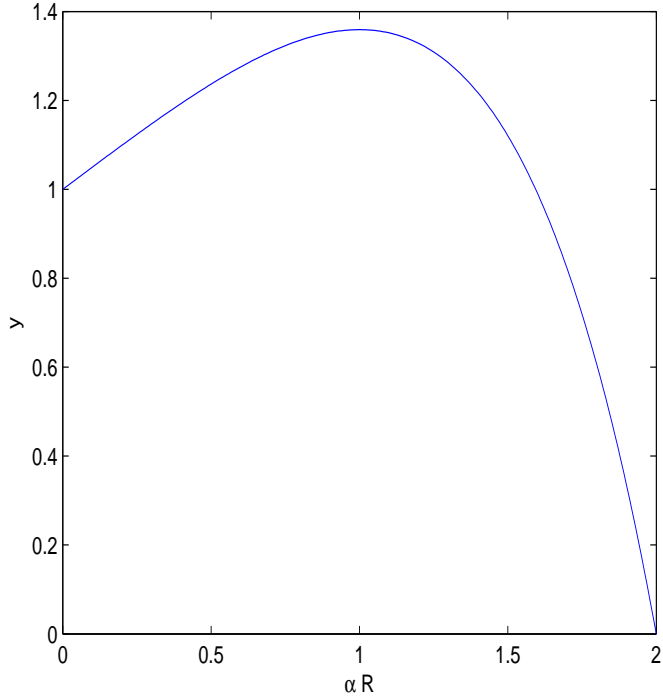}
\caption{\label{fig.4}The function $y=\left(1-\frac{\alpha R}{2}\right)\exp(\alpha R)$ versus $\alpha R$.}
\end{figure}

\begin{figure}[h]
\includegraphics[height=4.0in,width=4.0in]{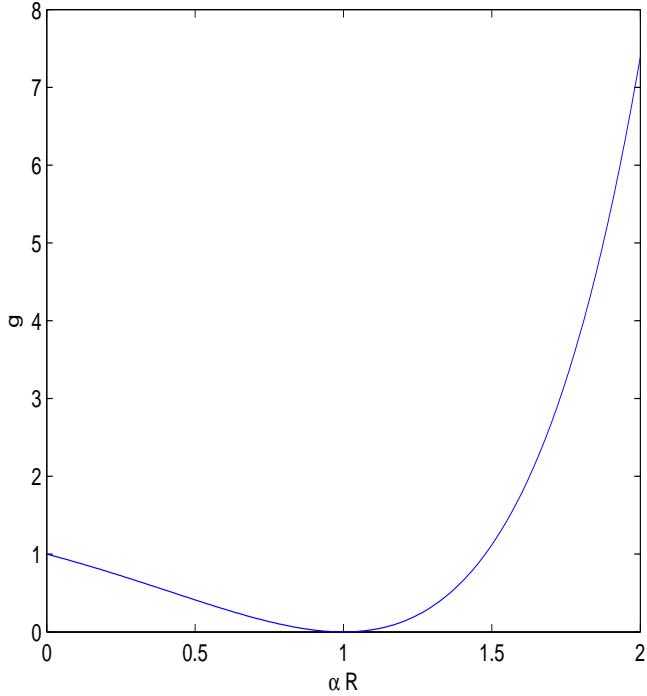}
\caption{\label{fig.5}The function $g=\left(1-\alpha R\right)^2\exp(\alpha R)$ versus $\alpha R$.}
\end{figure}

It should be noted that any viable inflationary model in F(R)-gravity has to be close to the Starobinsky model; that requires that the function $A(R)$ obeys the inequalities \cite{Ketov}
\begin{equation}
\mid A'(R)\mid < \frac{A(R)}{R},~~~~\mid A''(R)\mid < \frac{A(R)}{R^2},
\label{34}
\end{equation}
where
\begin{equation}
F(R)=R+R^2A(R).
\label{35}
\end{equation}
From Eqs.(1),(35) one obtains
\begin{equation}
A(R)=\frac{\exp(\alpha R)-1}{R}.
\label{36}
\end{equation}
By taking into account that $\alpha R\exp(\alpha R)>\exp(\alpha R)-1$ that is verified by Eq.3 for $\alpha R<1$, from Eqs.(34),(36), we find the restrictions on the value of $\alpha R$
\begin{equation}
\left(1-\frac{\alpha R}{2}\right)\exp(\alpha R)>1,~~~~\left(1-\alpha R\right)^2\exp(\alpha R)<1.
\label{37}
\end{equation}
The graph of the function (see Eq.(37)) $y=\left(1-\frac{\alpha R}{2}\right)\exp(\alpha R)$ is represented in Fig.\ref{fig.4}.
We note that nontrivial solution of the transcendental equation $\left(1-\alpha R/2\right)\exp(\alpha R)=1$ is $\alpha R=1.5936$. Thus, for $0<\alpha R<1.5936$ the first inequality in Eqs.(37) is valid.
The graph of the function (see Eq.(37)) $g=\left(1-\alpha R\right)^2\exp(\alpha R)$ is given by Fig.\ref{fig.5}.
It follows from Fig.5 that at $0<\alpha R<1$ the second inequality in Eqs.(37) is satisfied.
Next, any viable F(R)-gravity model of DE should be close to the standard $\Lambda$CDM-model. It gives the restriction on the function $F(R)$ ($R$ is positive in our metric) \cite{Ketov2}
\begin{equation}
\mid F(R)-\frac{1}{2}R\mid <R,~~\mid F'(R)-\frac{1}{2}\mid < 1,~~\mid RF''(R)\mid<1.
\label{38}
\end{equation}
The first inequality in (38) becomes $\alpha R<\ln 1.5=0.405$. The second inequality in (38) is equivalent to $(1+\alpha R)\exp(\alpha R)<1.5$. The graph of the function $f=\left(1+\alpha R\right)\exp(\alpha R)$ is given by Fig.6.
\begin{figure}[h]
\includegraphics[height=4.0in,width=4.0in]{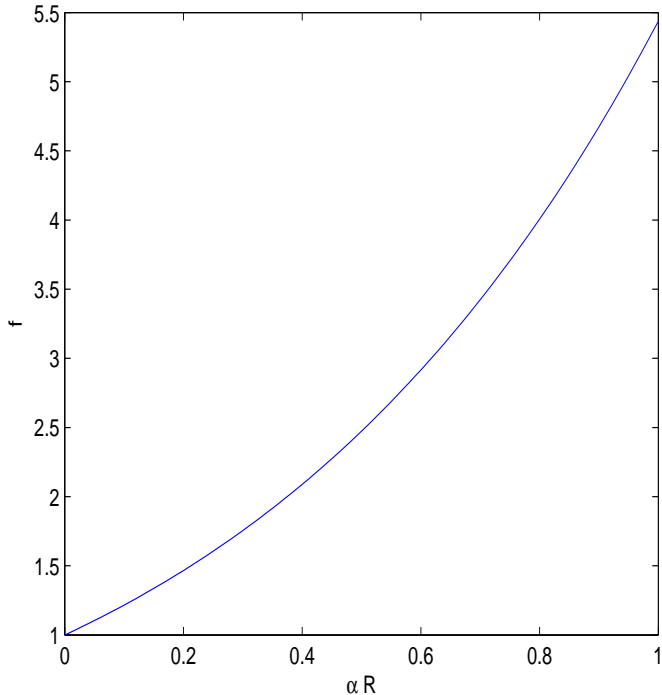}
\caption{\label{fig.6}The function $f=\left(1+\alpha R\right)\exp(\alpha R)$ versus $\alpha R$.}
\end{figure}
The equation $(1+\alpha R)\exp(\alpha R)=1.5$ has the solution $\alpha R=0.2127$. So, at $0<\alpha R<0.2127$ the second inequality in (38) is satisfied. The third inequality in (38) leads to $\alpha R\left(2+\alpha R\right)\exp(\alpha R)<1$. The plot of the function $z=\alpha R\left(2+\alpha R\right)\exp(\alpha R)$ is presented in Fig.7.
\begin{figure}[h]
\includegraphics[height=4.0in,width=4.0in]{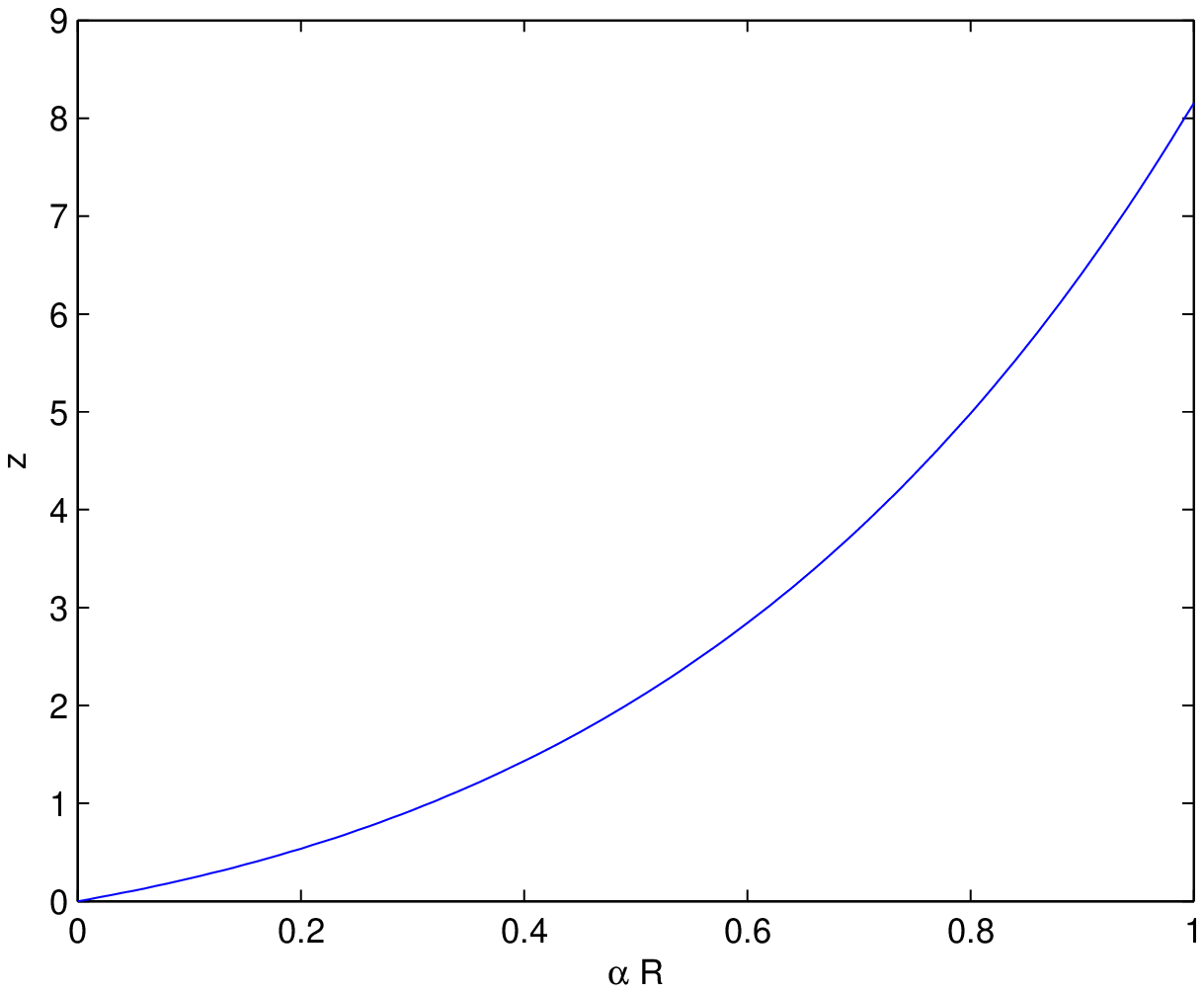}
\caption{\label{fig.7}The function $z=\alpha R\left(2+\alpha R\right)\exp(\alpha R)$ versus $\alpha R$.}
\end{figure}
The solution of the equation $\alpha R\left(2+\alpha R\right)\exp(\alpha R)=1$ is $\alpha R=0.3152$. So, at $0<\alpha R<0.3152$ the third inequality in (38) holds. To summarize, at $0<\alpha R<0.21$ all restrictions given by Eqs.(34),(38) are satisfied. Thus, the viability of the model based on Eq.(1) holds for $0<\alpha R<0.21$.

The slow-roll parameters read \cite{Liddle}
\begin{equation}
\varepsilon(\phi)=\frac{1}{2}M_{Pl}^2\left(\frac{V'(\phi)}{V(\phi)}\right)^2,~~~~\eta(\phi)=M_{Pl}^2\frac{V''(\phi)}{V(\phi)},
\label{39}
\end{equation}
were $M_{Pl}=\kappa^{-1}$ is the reduced Planck mass and primes are derivatives with respect to the field $\phi$. The conditions $\varepsilon(\phi)\ll 1$, $\mid\eta(\phi)\mid\ll 1$ are necessary for the slow-roll approximation.
From Eqs.(23)-(25) we obtain the slow-roll parameters expressed through the curvature
\begin{equation}
\varepsilon=\frac{\left(1-\alpha R\right)^2}{3\left(\alpha R\right)^2},~~~~
\eta=\frac{2\left(1-4\alpha R+(\alpha R)^3\right)}{3(\alpha R)^2\left(2+\alpha R\right)},
\label{40}
\end{equation}
where the dependance of curvature $R$ on the inflation field $\phi$ is given by Eq.(23) (see Fig.1). The plots of the functions $\varepsilon$, $\eta$ are represented in Fig.8 and Fig.9, respectively.
\begin{figure}[h]
\includegraphics[height=4.0in,width=4.0in]{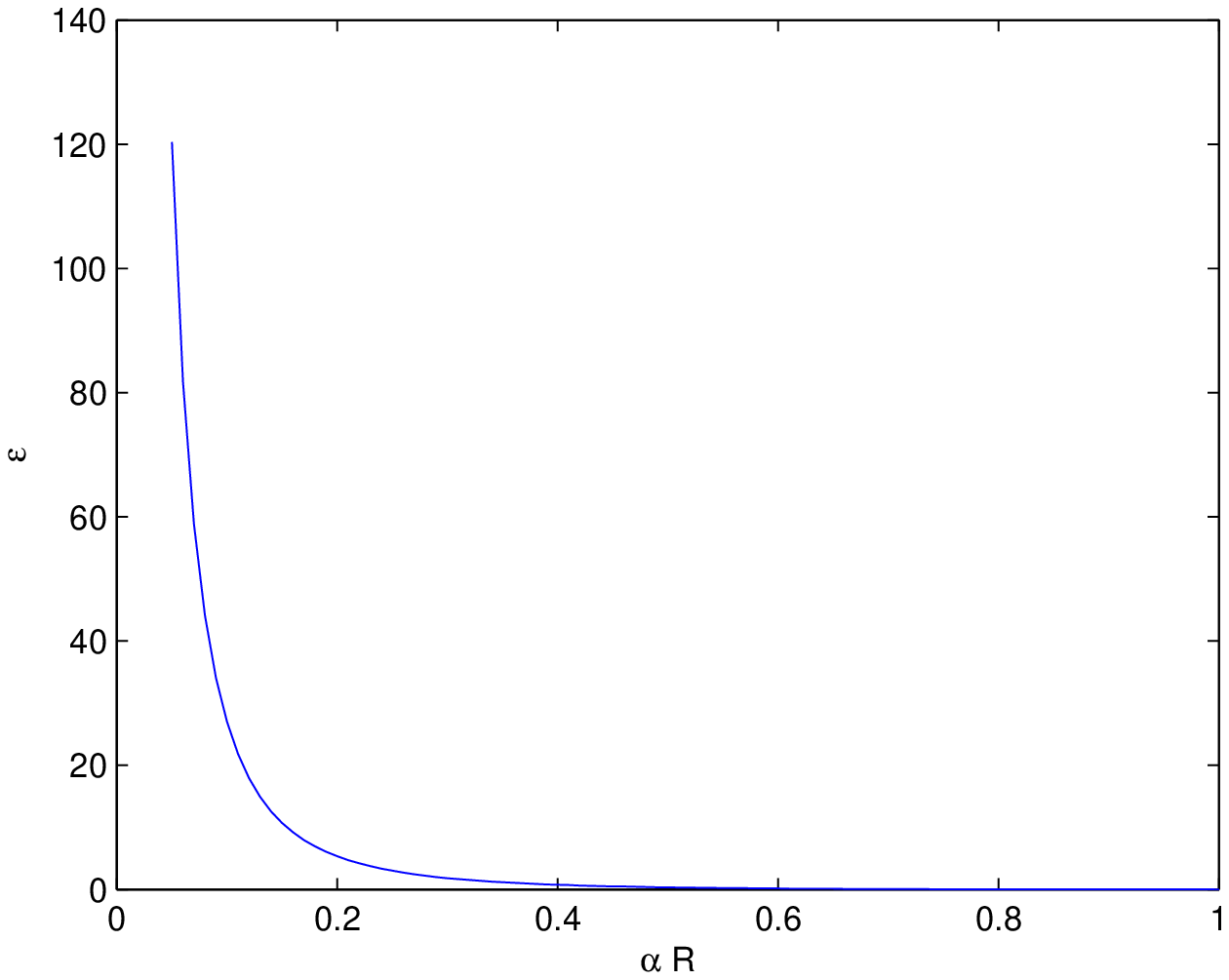}
\caption{\label{fig.8}The function $\varepsilon$ versus $\alpha R$.}
\end{figure}
\begin{figure}[h]
\includegraphics[height=4.0in,width=4.0in]{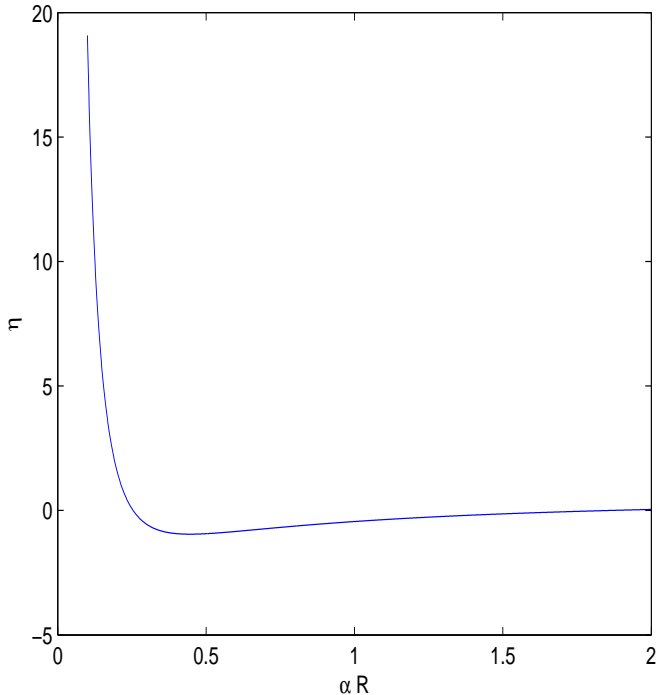}
\caption{\label{fig.9}The function $\eta$ versus $\alpha R$.}
\end{figure}
At $\alpha R>0.366$ the inequality $\varepsilon <1$ holds, and at $\alpha R>0.21$ we have $\mid\eta\mid <1$. It follows from Eqs.(40) that asymptotic of function $\varepsilon$ is $\lim_{\alpha R\rightarrow\infty}\varepsilon=1/3$ and $\lim_{\alpha R\rightarrow\infty}\eta=2/3$. As a result, the slow-roll approximation of the model ($\varepsilon(\phi)\ll 1$, $\mid\eta(\phi)\mid\ll 1$) is questionable.

One can calculate the age of the universe by evaluating the number of e-folds \cite{Ketov3}
\begin{equation}
N_e\approx \frac{1}{M_{Pl}^2}\int_{\phi_{end}}^{\phi}\frac{V(\phi)}{V'(\phi)}d\phi,
\label{41}
\end{equation}
where $\phi_{end}$ corresponds to the end of inflation. From Eqs.(23),(24),(41), after integration, we obtain e-folding number
\begin{equation}
N_e\approx 1.5\left[\alpha\left(R_{end}-R\right)+\ln\left(\frac{1-\alpha R_{end}}{1-\alpha R}\right)^{3/2}\left(\frac{1+\alpha R_{end}}{1+\alpha R}\right)^{1/2}\right].
\label{42}
\end{equation}
At $R_{end}=0$, $\alpha R=1-10^{-11}$, one finds from Eq.(42) the reasonable value $N_e\approx 55$. Thus, starting with $\alpha R\approx 1$ ($\kappa\phi\approx 2$) universe being in the unstable de Sitter space rolls down into the stable Minkowski space during the period characterized by $N_e$. The value of $N_e$ (Eq.(42)) is sensitive with the choice of starting high curvature regime because expression (42) possesses the singularity at $\alpha R$=1. Of course one can consider the example with the value $\alpha R=1-10^{-11}$ as a speculation. Therefore, the phenomenological gravity model considered may describe the evolution of the Universe in the low curvature regime.

The slope of the scalar power spectrum due to density perturbations is defined by
\begin{equation}
n_s=1-6\varepsilon+2\eta.
\label{43}
\end{equation}
From Eqs.(40),(43) we obtain
\begin{equation}
n_s=\frac{(\alpha R)^3+6(\alpha R)^2+2\alpha R-8}{3(\alpha R)^2(\alpha R+2)}.
\label{44}
\end{equation}
The tensor-to-scalar ratio is given by \cite{Liddle} $r=16\varepsilon$. The PLANCK data give \cite{Ade}
\begin{equation}
n_s=0.9603\pm 0.0073,~~~~r<0.11.
\label{45}
\end{equation}
It follows from Eq.44 that the experimental value of $n_s$ is not satisfied, and the inequality $r<0.11$ does not hold. We find from Eq.(44) that the maximum value of $n_s\approx 0.6$ is realized at high curvature $\alpha R\approx 2.4$. Thus, the model considered can give only approximate description of inflation.

\section{Conclusion}

A modified theory of gravity with exponential-like Lagrangian density and the fundamental length $L=\sqrt{\alpha}$ is considered and analyzed. We have investigated $F(R)$-model that admits de Sitter solutions without a cosmological constant. Therefore, possibly the cosmic acceleration arises from this theory of gravity and GR is only an approximation describing the Universe at the intermediate cosmic time. At $\alpha\rightarrow 0$ the action (2) approaches to the Einstein$-$Hilbert action. From the bound (4) obtained, we find the restriction on the fundamental length $L\leq 10^{-3}$ cm. We have found the static Schwarzschild$-$de Sitter solutions of the model and the potential of the scalar field in the scalar-tensor form of the theory. It was demonstrated that, for static solutions obtained, the de Sitter space is unstable and the Minkowski space with zero Ricci scalar is stable. Vacuum solutions are important in investigations of early and late time Universe. The model under consideration passes the matter stability test at $R_0<2/\alpha$. One can interpret this as a restriction on the size of the Universe at the beginning (the size of the Universe was greater than $L$). The slow-roll parameters $\varepsilon$, $\eta$ as well as the tensor-to-scalar ratio $n_s$ of the model have been calculated. We show that there is no slow-roll in the model considered. Therefore the model can describe the evolution of the Universe only in the low curvature regime. It follows from the analysis of the cosmological parameters that the inflation in this particular version of F(R)-gravity is possible but clearly not viable, because there is no slow roll and no agreement with the observed CMB data by the WMAP and PLANCK experiments.


\end{document}